\documentclass[useAMS,usenatbib]{mn2e}

\usepackage[T1]{fontenc}
\usepackage{pslatex}
\usepackage{amsmath}
\usepackage{amsfonts}
\usepackage{color}
\usepackage{url}
\usepackage{graphicx}
\usepackage{subfigure}
\usepackage{amssymb}
\usepackage{textcomp}
\usepackage{soul}
\usepackage{bm}
\usepackage{booktabs}
\usepackage{array}
\usepackage[draft]{hyperref}
 \graphicspath{{images/}}

{\newif\ifnotend
\notendtrue
\def\veclist{ABCDEFGHIJKLMNOPQRSTUVWXYZabcdefghijklmnopqrstuvwxyz.}
\def\top#1#2.{#1}
\def\tail#1#2.{#2.}
\loop\expandafter\xdef\csname v\expandafter\top\veclist\endcsname%
{{\noexpand\bf\expandafter\top\veclist}}
\edef\veclist{\expandafter\tail\veclist}
\if\veclist.\notendfalse\fi\ifnotend\repeat}

\def\pa{\partial}

\mathchardef\mhyphen="2D

\title[Thermal instability of rotating stratified atmospheres]{The effect of rotation on the thermal instability of stratified galactic atmospheres - I. Local analysis}

\author[Sobacchi \& Sormani]{Emanuele Sobacchi$^{1,2}$\thanks{Contact email: sobacchi@post.bgu.ac.il} and Mattia C. Sormani$^{3}$\thanks{Contact email: mattia.sormani@uni-heidelberg.de}\\
$^1$Physics Department, Ben-Gurion University, P.O.B. 653, Beer-Sheva 84105, Israel \\
$^2$Department of Natural Sciences, The Open University of Israel, 1 University Road, P.O.B. 808, Raanana 4353701, Israel \\
$^3$Universit\"{a}t Heidelberg, Zentrum f\"{u}r Astronomie, Institut f\"{u}r theoretische Astrophysik, Albert-Ueberle-Str. 2, 69120 Heidelberg, Germany}

\begin{document}

\date{}

\def\p{\partial}
\def\Omegap{\Omega_{\rm p}}

\newcommand{\di}{\mathrm{d}}
\newcommand{\bfx}{\mathbf{x}}
\newcommand{\bfe}{\mathbf{e}}
\newcommand{\bfxi}{\bm{\xi}}
\newcommand{\vlos}{\mathrm{v}_{\rm los}}
\newcommand{\Tspin}{T_{\rm s}}
\newcommand{\Tb}{T_{\rm b}}
\newcommand{\degree}{\ensuremath{^\circ}}
\newcommand{\Th}{T_{\rm h}}
\newcommand{\Tc}{T_{\rm c}}
\newcommand{\bfr}{\mathbf{r}}
\newcommand{\bfv}{\mathbf{v}}
\newcommand{\bfu}{\mathbf{u}}
\newcommand{\pc}{\,{\rm pc}}
\newcommand{\kpc}{\,{\rm kpc}}
\newcommand{\Myr}{\,{\rm Myr}}
\newcommand{\Gyr}{\,{\rm Gyr}}
\newcommand{\kms}{\,{\rm km\, s^{-1}}}
\newcommand{\de}[2]{\frac{\partial #1}{\partial {#2}}}
\newcommand{\cs}{c_{\rm s}}
\newcommand{\rb}{r_{\rm b}}
\newcommand{\rqu}{r_{\rm q}}
\newcommand{\nuP}{\nu_{\rm P}}
\newcommand{\thetaobs}{\theta_{\rm obs}}
\newcommand{\hatn}{\hat{\textbf{n}}}
\newcommand{\hatt}{\hat{\textbf{t}}}
\newcommand{\hatx}{\hat{\textbf{x}}}
\newcommand{\haty}{\hat{\textbf{y}}}
\newcommand{\hatz}{\hat{\textbf{z}}}
\newcommand{\hatX}{\hat{\textbf{X}}}
\newcommand{\hatY}{\hat{\textbf{Y}}}
\newcommand{\hatZ}{\hat{\textbf{Z}}}
\newcommand{\hatN}{\hat{\textbf{N}}}
\newcommand{\hater}{\hat{\mathbf{e}}_r}
\newcommand{\hateR}{\hat{\mathbf{e}}_R}
\newcommand{\hatephi}{\hat{\mathbf{e}}_\phi}
\newcommand{\hatez}{\hat{\mathbf{e}}_z}
\newcommand{\hateP}{\hat{\mathbf{e}}_P}
\newcommand{\hatePhi}{\hat{\mathbf{e}}_\Phi}
\newcommand{\hatetheta}{\hat{\mathbf{e}}_\theta}
\newcommand{\hatemu}{\hat{\mathbf{e}}_\mu}
\newcommand{\hatenu}{\hat{\mathbf{e}}_\nu}
\newcommand{\hatePL}{\hat{\mathbf{e}}_{P\Lambda}}
\newcommand{\nablaPL}{\nabla_{P\Lambda}}

\maketitle

\begin{abstract}
Observations show that (i) multiple gas phases can coexist in the atmospheres of galaxies and clusters; (ii) these atmospheres may be significantly rotating in the inner parts, with typical velocities that approach or even exceed the local sound speed. The thermal instability is a natural candidate to explain the formation of cold structures via condensation of a hotter gas phase. Here we systematically study the effect of rotation on the thermal stability of stratified plane-parallel atmospheres, using both analytical arguments and numerical simulations. We find that the formation of cold structures starting from small isobaric perturbations is enhanced in the regions where the rotation of the system is dynamically important (i.e. when the rotational velocity becomes comparable to the sound speed). In particular, the threshold value of the ratio between the cooling and dynamical time $t_{\rm cool}/t_{\rm dyn}$ below which condensations can form is increased by a factor up to $\sim 10$ in the presence of significant rotation. We briefly discuss the implications of our results for galaxies and clusters.
\end{abstract}


\section{Introduction}

The properties of galactic atmospheres are now reasonably constrained, both in the circumgalactic medium of galaxies \citep[e.g.][]{Putman+2012, Tumlinson+2017} and in the hot intracluster medium of galaxy clusters \citep[e.g.][]{Vikhlinin2006, Sanderson2006, Sanderson2009}. The hot gas phase presumably extends at least out to the virial radius, and has a typical temperature of $\sim 10^6{\rm\; K}$ in galaxies and of $\sim 10^7\mhyphen 10^8{\rm\; K}$ in clusters.

There is robust evidence that multiple gas phases can coexist in such galactic atmospheres. Observations of metal absorption lines in the spectra of background quasars show that low ionisation species, whose abundance peaks at temperatures of  $\sim 10^4 \mhyphen 10^5{\rm\; K}$, may be present out to the virial radius in the circumgalactic atmospheres of external galaxies \citep[e.g.][]{Werk2013, Werk2014}. The High Velocity Clouds (HVCs) around the Milky Way disc are probably the most spectacular example of a cold gas phase immersed in the hot gas of the circumgalactic medium. These $\kpc$ scale clouds are primarily detected through their HI $21{\rm cm}$ line emission, with the typical velocity dispersion suggesting temperatures of the order of $10^4{\rm\; K}$ \citep[e.g.][and references therein]{Putman+2012,Westmeier2018}. In the cool cores of galaxy clusters, filamentary structures extending tens of $\kpc$ from the centre of the potential have been directly imaged by H$\alpha$ emission \citep[e.g.][]{Conselice+2001, Fabian2008, McDonald2010, McDonald2011} and molecular line emission \citep[e.g.][]{SalomeCombes2004, Salome+2006}.

Thermal instability is a natural candidate to explain the formation of multi-phase structures in these galactic atmospheres. Recent studies have suggested that the condensation of linear perturbations in non-rotating stratified atmospheres requires that (i) the ratio of the cooling to the dynamical time $t_{\rm cool}/t_{\rm dyn}$ is less than a threshold value;\footnote{\citet{McCourt+2012} found a somewhat more stringent criterion, $t_{\rm cool}/t_{\rm dyn}\lesssim 1$, than \cite{Sharma+2012}, who found $t_{\rm cool}/t_{\rm dyn}\lesssim 10$. Previous claims attributed the difference to the different geometry (plane parallel vs spherical), but \citet{ChoudhurySharma2016} has shown that the difference is due to the fact that \citet{McCourt+2012} looked at the condensation at the location of $\min(t_{\rm cool}/t_{\rm ff})$, while \citet{Sharma+2012} considered condensations anywhere in the box, which happens further inside the location of the initial $\min(t_{\rm cool}/t_{\rm ff})$.} (ii) the feedback heating balances the cooling on large scales in order to prevent a cooling catastrophe \citep[e.g.][]{McCourt+2012, Sharma+2012}. However, the idea that this simple criterion regulates the thermal stability of hot galactic atmospheres may be an oversimplification \citep[see e.g.][and references therein]{Werner+2019}.

Independent lines of research suggest that galactic atmospheres are significantly rotating. For example, \citet{HodgesKluck+2016} measured the Doppler shifts of the OVII absorption line towards an ensemble of background quasars, and concluded that the circumgalactic medium of the Milky Way rotates with velocities of the order of $\sim 180\kms$. Note that since these measurements are only sensitive to the gas relatively close to the Galactic disc, this only implies that rotation is important out to radii of few tens of kpc, beyond which the Galactic corona is not expected to rotate significantly. Cosmological simulations also suggest that galactic coronae must rotate in the inner parts \citep[e.g.][]{Oppenheimer2018}. In particular, this study suggested that galactic coronae may become rotationally instead of pressure supported at distances $\lesssim \text{few tens of }\kpc$ from the galaxy. Thus both observations and theory suggest that galactic coronae rotate significantly in the inner parts, although the rotation is probably important only at radii $\ll$ than the virial radius.

The rotation of the intracluster medium is usually considered to be negligible. However, recent ALMA observations of a sample of 15 clusters showed that in 2 cases (Abell 262 and Hydra-A) the molecular gas has settled into a rotating disc \citep{Olivares+2019, Russell+2019}. Also note that in some clusters the rotation pattern may be hidden because the system is viewed face on, as suggested in the case of Abell 1835 by \citet{McNamara+2014}. It has also been suggested that a significant fraction ($\sim 25 \%$) of clusters may rotate coherently on ${\rm Mpc}$ scales with typical velocities of a few hundreds of $\kms$ (\citealt{ManolopoulouPlionis2017}; see also \citealt{Bianconi+2013}). Finally, the presence of cool core cold fronts, i.e. concentric or spiral density and temperature discontinuities in otherwise relaxed clusters \citep[see e.g.][]{MarkevitchVikhlinin2007}, may be associated with tangential flows with velocities of the order of the local sound speed \citep{Keshet+2010}. Hence, it is interesting to consider the possible impact of rotation on the condensation of cold gas also in the context of the intracluster medium.

\begin{figure}
\centering
\includegraphics[width=0.48\textwidth]{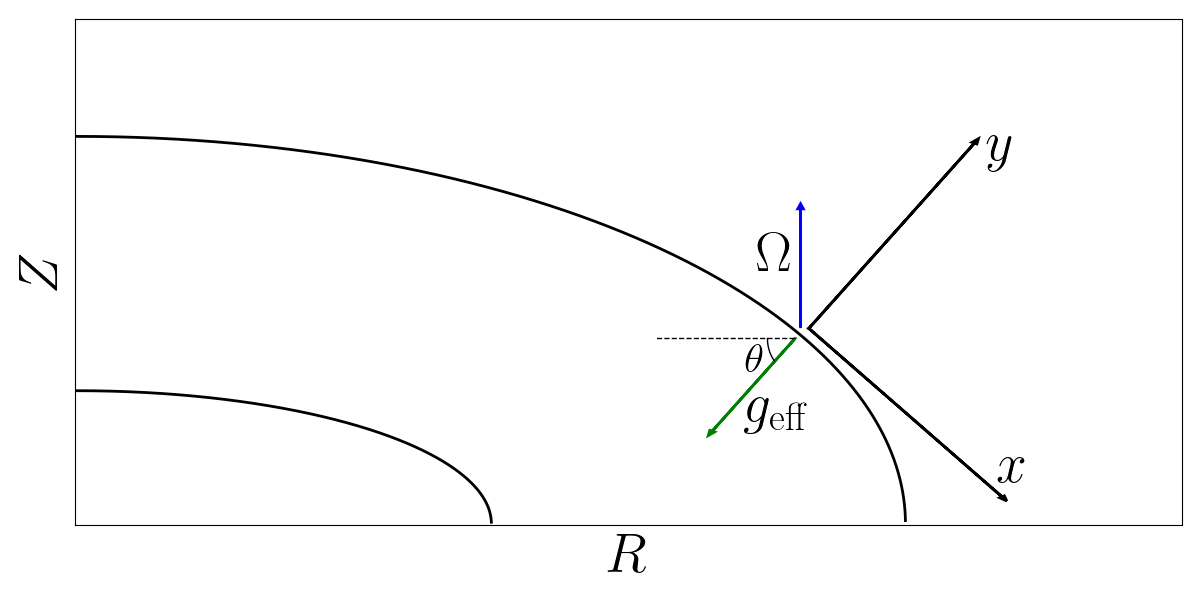}
\caption{Sketch (not to scale) of the local reference frame $xyz$ used in this paper (the $z$ axis is perpendicular to the $xy$ plane, along the $-\phi$ direction of the global cylindrical coordinates). The isobaric surfaces of the rotating atmosphere are shown with the black solid lines in the global $RZ$ reference frame. The equatorial plane is directed along the $R$ axis, while the angular velocity, $\Omega$, is directed along the $Z$ axis. The $y$ axis of the local frame is directed along the effective gravity $g_{\rm eff}$, which is perpendicular to the local isobaric surface and forms an angle $\theta$ with the $R$ axis.}
\label{fig:sketch}
\end{figure}

The thermal stability of rotating galactic atmospheres has been scarcely considered in the literature. The most systematic studies are probably the ones by \citet{Nipoti2010,NipotiPosti2014} and \citet{Nipoti+2015}, who studied the thermal stability of galactic coronae in the linear regime and found that the presence of rotation changes the dispersion relation in a non trivial way (namely, a new unstable mode appears). However, these analysis were limited to the linear regime, which is known to have serious limitations in the study of multiphase gas \citep{McCourt+2012}.

Since it is difficult to draw robust conclusions on the saturation of the instability based just on a linear analysis, we were motivated to carry out a study that clarifies the impact of rotation on the thermal stability of stratified galactic atmospheres. We study the evolution of a simplified plane-parallel stratified atmosphere in the presence of cooling and heating using both analytical arguments and numerical simulations. We include in the equations of motions the presence of rotation ($\Omega\neq0$), but we neglect the effects of differential rotation ($\nabla \Omega=0$). Our setup constitutes a generalisation of the one of \cite{McCourt+2012}, to which we have added the presence of rotation. This simplified problem is meant to represent the local conditions of either the intracluster or circumgalactic medium. Our local analysis is complemented by a companion study \citep[][hereafter Paper {\sc II}]{SormaniSobacchi2019} in which we perform global simulations of the Galactic corona and examine whether the effects of rotation on the thermal instability can lead to condensations resembling HVCs.

The paper is organised as follows. In Section \ref{sec:model} we describe our model. In Section \ref{sec:DR} we study the linear stability of our model. In Section \ref{sec:simulations} we use numerical simulations to study the development of the thermal instability in the non-linear regime. In Section \ref{sec:relevance} we briefly describe how our model can be used to interpret the formation of cold structures in the hot atmospheres of galaxies. We sum up in Section \ref{sec:conclusions}.

\section{Physical model}
\label{sec:model}

We are interested in the effects of rotation on the thermal stability of a rotating, stratified atmosphere in nearly equilibrium. This may be either the hot intracluster medium or the circumgalactic medium of an individual galaxy. We use a simplified problem of a stratified, plane parallel atmosphere, which is governed by the equations of fluid dynamics:
\begin{align}
\label{eq:continuity}
& \frac{\pa\rho}{\pa t} + \nabla\cdot\left(\rho{\bf v}\right)=0\;, \\
\label{eq:euler}
& \left[\frac{\pa}{\pa t} + \left({\bf v}\cdot\nabla\right)\right]{\bf v} = -\frac{\nabla P}{\rho} -2{\pmb \Omega}\times{\bf v} +{\bf g}_{\rm eff}\;,\\
\label{eq:entropy}
& \frac{P}{\gamma -1}\left[\frac{\pa}{\pa t} + \left({\bf v}\cdot\nabla\right)\right]\sigma = \mathcal{H-L}\;,
\end{align}
where $\rho$ is the mass density, ${\bf v} = (v_x,v_y,v_z)$ is the velocity in the rotating frame, $P$ is the pressure and $\sigma=\log\left(P/\rho^\gamma\right)$ is the entropy. The term $-2{\pmb \Omega}\times{\bf v}$ on the right hand side of Eq. \eqref{eq:euler} is a Coriolis term and represents the effect of rotation, where ${\pmb \Omega}=\Omega\left(-\cos\theta{\bf e}_x + \sin\theta{\bf e}_y\right)$ is the angular velocity of the rotating frame. In our problem $\Omega$ and $\theta$ are constants so our setup neglects the effect of shear. The last term on the right hand side of Eq. \eqref{eq:euler} represents an external effective gravity term where ${\bf g}_{\rm eff}= - g_{\rm eff}(y){\bf e}_y$ and $g_{\rm eff}>0$. Fig. \ref{fig:sketch} schematically sketches the underlying physical picture: our simplified problem is meant to represent a local frame in a steady, stratified atmosphere rotating with a purely azimuthal velocity. The local frame is oriented such that the effective gravity ${\bf g}_{\rm eff}=-\nabla\Phi + \Omega^2{\bf R}$, which includes the centrifugal terms, is directed along the negative $y$ axis. Note that $\left|{\bf g}_{\rm eff}\right|$ decreases as the system becomes more rotationally supported. Also note that the condition that the atmosphere is in equilibrium implies that ${\bf g}_{\rm eff} = \nabla P/\rho$.

Equation \eqref{eq:entropy} is the energy equation. The terms on the right hand side account for the presence of heating and cooling respectively. Throughout this paper we assume that (i) the cooling function depends on the state of the gas, namely $\mathcal{L}=\mathcal{L}\left(\rho,T\right)$; (ii) the heating function depends only on the position, namely $\mathcal{H}=\mathcal{H}\left({\bf x}\right)$; (iii) when the system is in equilibrium, the cooling is exactly balanced by the heating. Though highly idealised, this choice allows one to isolate the local effect of the thermal instability, preventing the catastrophic collapse of the atmosphere that is occurring in the absence of any form of feedback. In this paper we adopt an adiabatic index $\gamma=5/3$, the value for monoatomic ideal gases, and an ideal equation of state $P=\left(\rho/\mu m_p\right)kT$, where $T$ is the temperature, $k$ is the Boltzmann constant, $\mu$ is the mean molecular weight, and $m_p$ is the proton mass.

Note that the set of Eqs. \eqref{eq:continuity}-\eqref{eq:entropy} is identical to the one studied by \citet{McCourt+2012} except for the addition of the Coriolis term, which accounts for the presence of rotation. 

\subsection{Equilibrium solution}
\label{sec:unperturbed}

The system of Eqs. \eqref{eq:continuity}-\eqref{eq:entropy} has stationary solutions ($\partial/\partial t=0$) in which the velocity vanishes everywhere (${\bf v}=0$). It is straightforward to check that under these assumptions Eq. \eqref{eq:continuity} is automatically satisfied, and that Eq. \eqref{eq:entropy} can be satisfied with an appropriate choice of the heating function. The only non-vanishing component of Eq. \eqref{eq:euler} is the one along $y$, which gives
\begin{equation}
\label{eq:euler_y}
\frac{{\rm d}P}{{\rm d}y}= - g_{\rm eff}\rho\;.
\end{equation}
We consider the isothermal solution of Eq. \eqref{eq:euler_y} such that $P=c_{\rm s}^2\rho$, where $c_{\rm s}$ is the isothermal sound speed. This immediately gives
\begin{equation}
\label{eq:steady_state}
\rho = \rho_0 \exp\left(-\int_0^y\frac{{\rm d}y}{H}\right)\;,
\end{equation}
where
\begin{equation}
H=\frac{c_{\rm s}^2}{g_{\rm eff}}
\end{equation}
is a natural length scale for the system. Note that $H$ depends on $y$ unless $g_{\rm eff}$ is assumed to be constant.

\subsection{Heating and cooling functions}

Following \citet{McCourt+2012}, we adopt a simple parametrisation for the cooling function, namely
\begin{equation}
\label{eq:coolL}
\mathcal{L}=\Lambda_0\left(\frac{\rho}{\rho_0}\right)^2\left(\frac{T}{T_0}\right)^{1/2}\;,
\end{equation}
which is appropriate for an optically thin plasma in which the cooling is dominated by thermal bremsstrahlung. The constant $\Lambda_0$ has units of ${\rm erg}\;{\rm cm}^{-3}\;{\rm s}^{-1}$. Once the equilibrium density and temperature are specified, the heating function is determined by the requirement that $\mathcal{H}\left({\bf x}\right)=\mathcal{L}\left(\rho,T\right)$, which prevents the catastrophic collapse of the atmosphere.

\subsection{Important time scales}
\label{sec:timescales}

The length scale $H$ is associated to the natural dynamical time scale
\begin{equation} \label{eq:tdyn}
t_{\rm dyn}=\sqrt{2}\frac{c_{\rm s}}{g_{\rm eff}}\;.
\end{equation}
The dynamical time becomes large in the limit $g_{\rm eff}\to 0$, which may happen in systems with substantial rotational support. Note that the dynamical time may be also expressed as $t_{\rm dyn}=\sqrt{2H/g_{\rm eff}}$, where $H=c_{\rm s}^2/g_{\rm eff}$, which for a non-rotating system coincides with the usually adopted definition of free-fall time. We prefer our alternative definition because it is based only on local quantities, and it is therefore easier to generalise to the rotating, non spherically symmetric atmospheres that we consider in Paper {\sc II}.

A second important time scale,
\begin{equation}
t_{\rm rot}=\frac{2\pi}{\Omega}\;,
\end{equation}
is associated with the angular velocity of the system. As we will see in Section \ref{sec:DR}, these time scales are respectively associated with the buoyant and epicyclic oscillations of a blob of gas around equilibrium. The system is essentially pressure supported in the regime $t_{\rm dyn}/t_{\rm rot}\ll 1$, while it becomes rotationally supported in the regime $t_{\rm dyn}/t_{\rm rot}\gg 1$.

Finally, the thermal time scale is given by the ratio of the internal energy density to the cooling function, which gives
\begin{equation} \label{eq:tcool}
t_{\rm cool}=\frac{1}{\gamma-1}\frac{P}{\mathcal{L}}\;.
\end{equation}
As we will see in Section \ref{sec:DR}, the cooling time scale regulates the growth rate of the thermally unstable modes.\footnote{In general there is a factor $\Delta(T) = 2 - \di \log \Lambda(T) / \di \log T$ between $t_{\rm cool}$ defined here and the growth rate of the thermal instability for a cooling function of the form $\mathcal{L} = (\rho/\rho_0)^2 \Lambda(T)$. For the idealised cooling function assumed in this paper, $\mathcal{L}\propto T^{1/2}$, this factor is $\Delta(T) = 3/2$ (compare Eqs. \ref{eq:tcool} and \ref{eq:omegath}).} Note that $t_{\rm cool}$ depends on $y$ in a stratified atmosphere.

\section{Linear stability analysis}
\label{sec:DR}

We linearise Eqs. \eqref{eq:continuity}-\eqref{eq:entropy} with Eulerian perturbations of the form $F+\delta F\exp\left(ik_x x + ik_y y + ik_z z -i\omega t\right)$, where $F$ is the equilibrium value and $\left|\delta F\right|\ll \left|F\right|$. We work in the WKB limit $kH\gg 1$, namely we assume the wavelength of the perturbation to be much shorter than the typical scale of the system. We use the Boussinesq approximation, which consists in (i) neglecting $\omega\delta\rho/\rho$ with respect to $k\delta v$ in the linearised continuity equation; (ii) neglecting $\delta P/P$ with respect to $\delta\rho/\rho$ in the linearised energy equation. This approximation allows one to study modes that are much slower than sound waves; we refer the reader to \citet{BalbusPotter2016} for a more detailed discussion.

In the approximation described above, the linearised continuity equation \eqref{eq:continuity} is
\begin{equation}
\label{eq:lin_cont}
{\bf k}\cdot\delta{\bf v}=0\;.
\end{equation}
The linearised Euler's equation \eqref{eq:euler} is
\begin{equation}
\label{eq:lin_euler}
-i\omega\delta{\bf v}= -i{\bf k}\frac{\delta P}{\rho} +{\bf g}_{\rm eff}\frac{\delta\rho}{\rho} -2{\pmb\Omega}\times\delta{\bf v}\;.
\end{equation}
In the Boussinesq approximation, the perturbation of the cooling function \eqref{eq:coolL} is $\delta\mathcal{L/L}=3/2\times\delta\rho/\rho$. Hence, the linearised energy equation \eqref{eq:entropy} gives
\begin{equation}
\label{eq:lin_entropy}
i\omega\gamma\frac{\delta\rho}{\rho} + \delta{\bf v}\cdot\nabla\sigma = -\frac{3}{2}\left(\gamma-1\right)\frac{\mathcal{L}}{P}\frac{\delta\rho}{\rho}\;.
\end{equation}

In order to get the dispersion relation, it is useful to take the cross product of Eq. \eqref{eq:lin_euler} with ${\bf k}$. Using Eq. \eqref{eq:lin_cont} to eliminate ${\bf k}\cdot\delta{\bf v}$, we find
\begin{equation}
\label{eq:lin_euler_bis}
-i\omega {\bf k}\times\delta{\bf v} =\left({\bf k}\times{\bf g}_{\rm eff}\right)\frac{\delta\rho}{\rho} + 2\left({\bf k}\cdot{\pmb\Omega}\right)\delta{\bf v}\;.
\end{equation}
Taking again the cross product of Eq. \eqref{eq:lin_euler_bis} with ${\bf k}$ and then using Eq. \eqref{eq:lin_euler_bis} to express ${\bf k}\times\delta{\bf v}$, we find
\begin{align}
\label{eq:lin_euler_ter}
i\omega k^2 & \left[1-4\frac{\left({\bf k}\cdot{\pmb\Omega}\right)^2}{k^2\omega^2}\right]\delta{\bf v} = \nonumber\\
& = \left[{\bf k}\times\left({\bf k}\times{\bf g}_{\rm eff}\right) + \frac{2i}{\omega}\left({\bf k}\cdot{\pmb\Omega}\right)\left({\bf k}\times{\bf g}_{\rm eff}\right)\right]\frac{\delta\rho}{\rho} \;.
\end{align}
Taking the dot product of Eq. \eqref{eq:lin_euler_ter} with $\nabla\sigma$ and using Eq. \eqref{eq:lin_entropy},
we finally find the following dispersion relation:
\begin{equation}
\label{eq:DR}
\omega^3 +i\omega_{\rm th}\omega^2 -\left(\omega_{\rm BV}^2+\omega_{\rm rot}^2\right)\omega -i\omega_{\rm rot}^2\omega_{\rm th}=0\;,
\end{equation}
where we have defined
\begin{align}
\omega_{\rm BV} & = \frac{g_{\rm eff}}{c_{\rm s}}\sqrt{\frac{\gamma-1}{\gamma}\left(1-\frac{k_y^2}{k^2}\right)}\\
\omega_{\rm rot} & = 2\frac{\left|{\bf k}\cdot{\pmb\Omega}\right|}{k} \\
\omega_{\rm th} & = -\frac{3}{2}\frac{\gamma-1}{\gamma}\frac{\mathcal{L}}{P}\;. \label{eq:omegath}
\end{align}
The dispersion relation \eqref{eq:DR} is formally identical to the one derived by \citet{Nipoti2010}, which reassures us that our idealised local setup is able to capture the essential physical ingredients of the full problem. Note that $|\omega_{\rm BV} t_{\rm dyn}|$ and $|\omega_{\rm rot} t_{\rm rot}|$ depend on the direction of $\mathbf{k}$, while $| \omega_{\rm th} t_{\rm cool}|$ does not.

\subsection{Some comments on the dispersion relation}

In the relevant case $\omega_{\rm th}<0$,\footnote{In the case $\omega_{\rm th}>0$ a thermal instability such that of \citet{Field1965} is not possible.} the dispersion relation \eqref{eq:DR} has always one or more solutions such that ${\rm Im}\left(\omega\right)>0$. Thus, being the system always formally unstable, the important question is whether the instability actually leads to the condensation of overdense clumps or not. Though the investigation of the instability in the non-linear regime clearly requires numerical simulations, studying the proper modes gives some insights on what is likely to happen.

A point that draws attention onto the importance of the non-linear effects is the following. If the atmosphere is isentropic (instead of isothermal as in our setup), the Brunt-V\"{a}is\"{a}l\"{a} frequency $\omega_{\rm BV}$ vanishes and the dispersion relation for a non-rotating corona is identical to the classical one of \citet{Field1965} for an infinite homogeneous gas. Thus, according to the linear analysis, an isentropic atmosphere should be very unstable and lead to condensations. However, it turns out that this is not the case: gravity is still able to suppress the instability, which saturates in a manner similar to the isothermal case \citep{McCourt+2012, Choudhury+2019}. While this highlights the limitations of the linear analysis, the results of the linear theory nevertheless constitute a useful guide to interpret the numerical simulations that we present in Section \ref{sec:simulations}.

The dispersion relation becomes more transparent in two limiting cases, namely (i) a fast cooling regime, which occurs when $t_{\rm cool}\ll t_{\rm dyn}$; (ii) a slow cooling regime, which occurs when $t_{\rm cool}\gg t_{\rm dyn}$. Note that the effective gravity $g_{\rm eff}$ is smaller in rotationally supported systems compared to purely pressure supported ones for the same value of $g$ (and therefore for the same value of $t_{\rm cool}/t_{\rm ff}$, where $t_{\rm ff}$ is calculated using $g$ and not $g_{\rm eff}$ in Eq. \ref{eq:tdyn}). Since the quantity that enters the dispersion relation is $g_{\rm eff}$, not $g$, this means that regions of stratified galactic atmospheres where the rotation is dynamically important fall more easily in the fast cooling regime. What matters is not the external gravity, but the external gravity diminished by the local centrifugal acceleration felt by rotating fluid elements. Hence condensation is more likely to happen in a rotating system even without the effect that we discovered in this paper, namely that the production of condensation is enhanced even when $g_{\rm eff}$ is kept fixed due to the presence of the Coriolis term in the equation of motion (see Fig. \ref{fig:condensation}). 

\subsubsection{Fast cooling regime $\left(t_{\rm cool}\ll t_{\rm dyn}\right)$} \label{sec:fastcooling}

Since in this case the Brunt-V\"{a}is\"{a}l\"{a} frequency is necessarily small ($\omega_{\rm BV}\ll\left|\omega_{\rm th}\right|$), we look for a solution of the dispersion relation of the form $\omega=\sum_{i=0}^{\infty} a_i\omega_{\rm BV}^i$. The solution of Eq. \eqref{eq:DR}, approximated with the first terms of the expansion, gives
\begin{equation}
\label{eq:mode1}
\omega=-i\omega_{\rm th} +i\frac{\omega_{\rm BV}^2}{\omega_{\rm th}^2+\omega_{\rm rot}^2}\omega_{\rm th}
\end{equation}
and
\begin{equation}
\label{eq:mode2}
\omega=\pm\omega_{\rm rot} \pm\frac{1}{2}\frac{\omega_{\rm BV}^2}{\omega_{\rm th}^2+\omega_{\rm rot}^2}\omega_{\rm rot} -\frac{i}{2}\frac{\omega_{\rm BV}^2}{\omega_{\rm th}^2+\omega_{\rm rot}^2}\omega_{\rm th}\;.
\end{equation}
The first corrections to the leading terms are of the order of $\omega_{\rm BV}^2/\omega_{\rm th}^2\ll 1$. Also note that here we have used only the fact that $\omega_{\rm BV}\ll\left|\omega_{\rm th}\right|$. Hence, Eqs. \eqref{eq:mode1} and \eqref{eq:mode2} are valid both when the rotation is fast ($t_{\rm rot}\ll t_{\rm cool}$) or slow ($t_{\rm rot}\gg t_{\rm cool}$) with respect to the cooling time.

Since $\omega_{\rm th}<0$, the mode \eqref{eq:mode1} corresponds to a thermally unstable condensation. Apart from a small correction due to buoyancy, the dispersion relation is analogous to the classical result of \citet{Field1965} for the case of a homogeneous, isotropic medium. Physically, one can argue that in the limit $t_{\rm cool}\ll t_{\rm dyn}$ the thermal condensation of an overdense clump occurs before gravity significantly displaces the clump from its initial position. Hence, in the fast cooling regime condensations are always expected to form due to the presence of this mode.

The mode described by Eq. \eqref{eq:mode2} instead essentially describes an epicyclic oscillation. The oscillation is overstable (i.e., its amplitude slowly grows over the course of many oscillations, but not monotonically) due to the effect of cooling. Overstable perturbations typically do not have enough time to cool before being disrupted by other processes, such as for example the Kelvin-Helmholtz instability \citep[e.g.][]{Joung+2012}.

\subsubsection{Slow cooling regime $\left(t_{\rm cool}\gg t_{\rm dyn}\right)$}
\label{sec:slowcooling}

In the slow cooling regime the situation becomes more intricate. Here we present some analytic arguments suggesting that thermal condensation may develop in this regime only if rotation is dynamically important.

If the thermal frequency is small with respect to the Brunt-V\"{a}is\"{a}l\"{a} frequency ($\left|\omega_{\rm th}\right|\ll\omega_{\rm BV}$), which is always the case unless ${\bf k}$ is almost parallel to direction of the local gravity field, we may look for a solution of the dispersion relation of the form $\omega=\sum_{i=0}^{\infty} a_i\omega_{\rm th}^i$. The first terms in the expansion are
\begin{equation}
\label{eq:mode3}
\omega = -i\frac{\omega_{\rm rot}^2}{\omega_{\rm BV}^2 + \omega_{\rm rot}^2}\omega_{\rm th}\;.
\end{equation}
and
\begin{equation}
\label{eq:mode4}
\omega=\pm \sqrt{\omega_{\rm BV}^2 + \omega_{\rm rot}^2} -\frac{i}{2}\frac{\omega_{\rm BV}^2}{\omega_{\rm BV}^2 + \omega_{\rm rot}^2} \omega_{\rm th}
\end{equation}

Eq. \eqref{eq:mode3} describes a purely unstable mode, which may lead to the condensation of dense clumps. Interestingly, the growth rate of the instability vanishes if $\Omega=0$. The fact that an unstable mode appears as the result of rotation was already noticed by \citet{Nipoti2010}. The growth rate of the instability approaches the classical \citet{Field1965} result in the limit $\omega_{\rm rot}\gg\omega_{\rm BV}$. This suggests that condensation may occur even in the case $\left|\omega_{\rm th}\right|\ll\omega_{\rm BV}$ if the effect of rotation is dynamically important, which gives a strong motivation to study the thermal instability of rotating atmospheres in the non-linear regime. The physical reason why this new mode appears in the presence of rotation is that when the buoyancy is slow with respect to rotation (but not with respect to cooling), an overdense clump may condense without undergoing a buoyant oscillation if the excess gravity is balanced by the Coriolis force (see appendix of Paper {\sc II} for a more detailed discussion).

The mode \eqref{eq:mode4} essentially describes a buoyant oscillation, which turns out to be overstable due to the effect of cooling, similarly to the non-rotating case \citep{Malagoli+1987,Binney+2009}. As discussed in Sect. \ref{sec:fastcooling}, it seems unlikely that such an overstability can result in a genuine thermal condensation, because an overdense clump sinks, mixes with the surrounding gas and is disrupted by other processes such as the Kelvin-Helmholtz instability before having time to cool significantly.

If instead ${\bf k}$ is nearly parallel to direction of the local gravity field, we have $\omega_{\rm BV}\ll\left|\omega_{\rm th}\right|$, and thus we recover the modes \eqref{eq:mode1}-\eqref{eq:mode2}. Indeed, setting $k^2\sim k_y^2$ one finds $\omega_{\rm BV}\sim 0$.\footnote{This is due to the fact that in this case the displacement $\delta{\bf v}$ is perpendicular to ${\bf g}_{\rm eff}$ (see Eq. \ref{eq:lin_cont}), so a displaced overdense clump does not feel any restoring buoyant force.} Thus according to the linear analysis the system is formally unstable. However, it has been argued that this does not lead to condensations because of non-linear effects that set in before condensations can occur \citep{Malagoli+1987,Binney+2009}: as soon as a blob is slightly denser than its surroundings (but still not dense enough to be called a proper condensation), it will start sinking, thus `changing' the direction of ${\bf k}$ which cannot remain parallel to ${\bf g}_{\rm eff}$ for a time sufficiently long to lead to a condensation (as discussed above, modes in which ${\bf k}$ is not parallel to ${\bf g}_{\rm eff}$ are not unstable). This behaviour has been confirmed by numerical simulations \citep[e.g.][]{McCourt+2012}. In Section \ref{sec:simulations} we argue that this conclusion may be relaxed if the rotation of the system becomes dynamically important because sinking of an overdense blob is prevented by the conservation of angular momentum.

\section{Numerical experiments}
\label{sec:simulations}

\begin{figure*}
\centering
\includegraphics[width=0.95\textwidth]{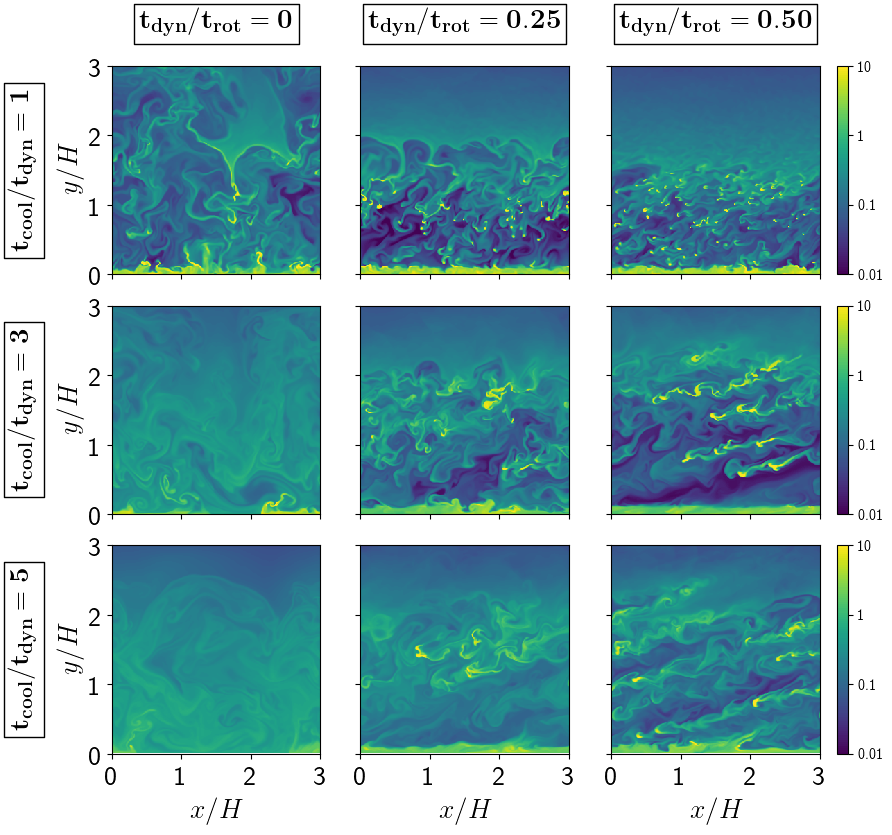}
\caption{Density field $\rho/\rho_0$ at $t=7 t_{\rm cool}$ of our simulations for different values of the ratios between the relevant timescales. From left to right: $t_{\rm dyn}/t_{\rm rot}=0, 0.25, 0.5$. From top to bottom: $t_{\rm cool}/t_{\rm dyn}=1,3,5$. The only difference between the left column and all the other panels is the presence of the Coriolis force in the latter (see term $2{\pmb \Omega}\times {\bf v}$ in Eq. \ref{eq:euler}). In the non rotating case $t_{\rm dyn}/t_{\rm rot}=0$ (left column) our results are consistent with those of \citet{McCourt+2012}. The presence of rotation (central and right columns) enhances the formation of condensations. The direction of $\pmb \Omega$ is such that $\theta=15 \degree$ in all panels. As discussed in Section \ref{sec:timescales}, our definition of dynamical time coincides with the free-fall time for a non-rotating atmosphere and is its natural generalisation in the case of a rotating atmosphere.}
\label{fig:main}
\end{figure*}

\begin{figure*}
\centering
\includegraphics[width=0.94\textwidth]{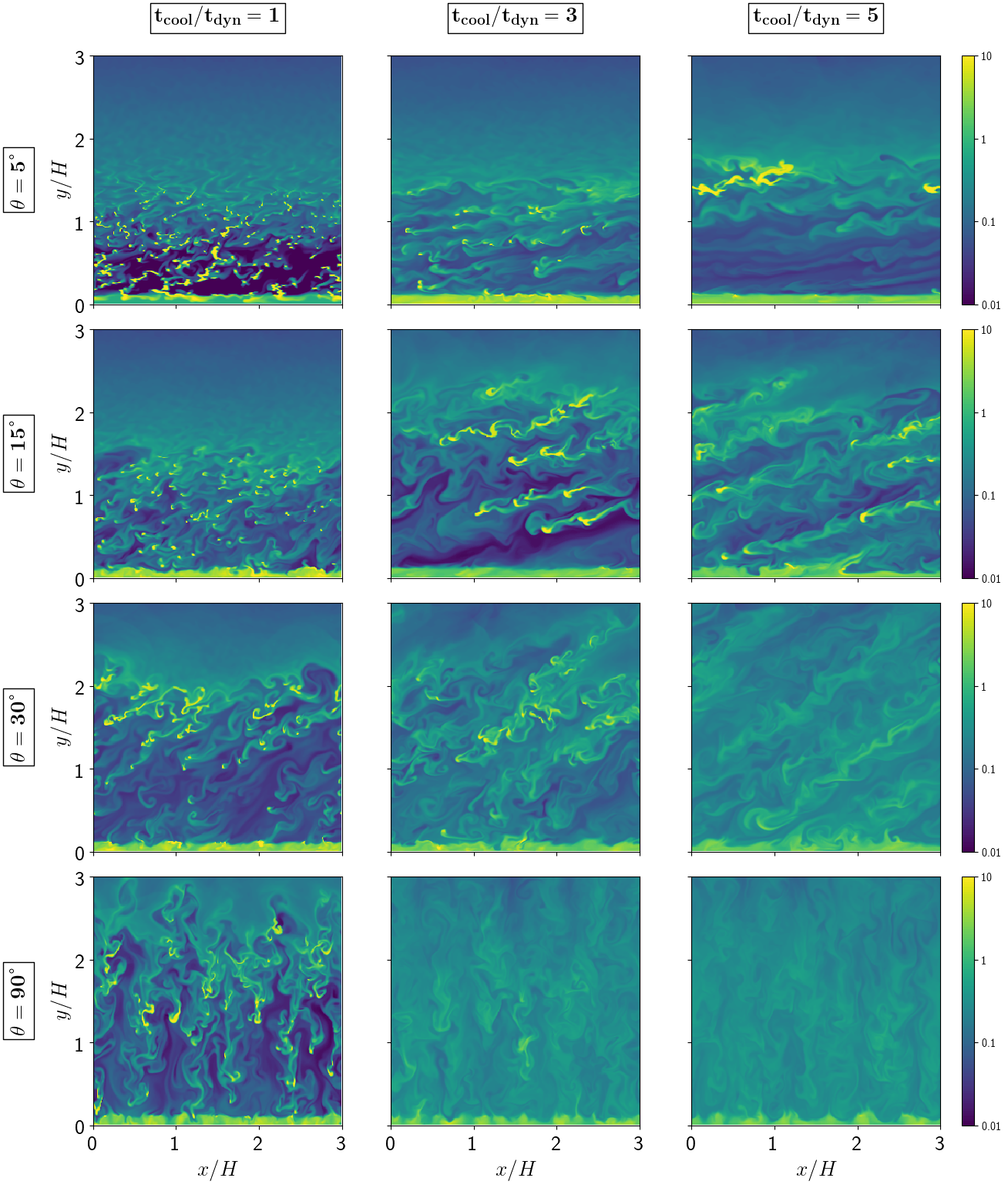}
\caption{Density field $\rho/\rho_0$ at $t=7 t_{\rm cool}$ of our simulations for various values of $\theta$ and $t_{\rm cool}/t_{\rm dyn}$ (see Fig. \ref{fig:sketch} for the definition of $\theta$). From top to bottom: $\theta=5, 15, 30, 90 \degree$. From left tor right: $t_{\rm cool}/t_{\rm dyn}=1,3,5$. All the simulations in this figure have $t_{\rm dyn}/t_{\rm rot}=0.5$. Note that the second row coincides with the right column of Fig. \ref{fig:main}. The slanted structures seen in the cold gas are due to the fact that gas tends to move in the direction parallel to $\pmb \Omega$ due to the conservation of angular momentum.}
\label{fig:angle}
\end{figure*}

\subsection{Simulation setup}

We study the non-linear development of the instability using the publicly available hydrodynamic code {\small PLUTO} version 4.3 \citep{Pluto2007}. Following \citet{McCourt+2012}, we take
\begin{equation}
g_{\rm eff}=g_0\frac{y/y_0}{\left[1+\left(y/y_0\right)^2\right]^{1/2}}\;.
\end{equation}
Hence, $g_{\rm eff}$ is approximately constant for $y\gtrsim y_0$, while it smoothly decreases to $0$ when $y \to y_0$. We work in units $g_0=c_{\rm s}=\rho_0=T_0=1$. We set $y_0=0.1$, and we turn the cooling off when $y<y_0$. We furthermore set a temperature floor $T_{\rm floor}=1/20$. We have checked that our conclusions are unaffected by the choice of $T_{\rm floor}$ and $y_0$, provided that $y_0\ll 1$ and $T_{\rm floor}\ll 1$.

We run a set of 2.5D simulations, i.e. the simulations are on a two-dimensional $(x,y)$ grid but we allow the velocity to have three components $(v_x,v_y,v_z)$, which is equivalent to assume that the system is translationally invariant along the $z$ direction. The system evolves according to Eqs. \eqref{eq:continuity}-\eqref{eq:entropy}. Our simulation domain is $x \times y = [0, 3] \times [0,3]$, and we use a uniformly spaced Cartesian grid with $300 \times 300$ points. We use periodic boundary conditions in the $x$ direction, and reflective boundary conditions in the $y$ direction. As initial conditions we take the equilibrium state described in Section \ref{sec:model}, on which we add some random noise by applying density perturbations seeded with a flat spectrum in the range $2\pi/3<k<40\pi/3$. The perturbations have Gaussian random amplitudes with an rms value of $10^{-2}$ and are isobaric, so that $P(y)$ is the same as in the unperturbed equilibrium state. Also the velocity is $\bfv=0$ initially, as in the unperturbed equilibrium state. We use the following parameters: {\sc rk2} time-stepping, no dimensional splitting, {\sc roe} Riemann solver and the {\sc mc} flux limiter.

We explore all the possible combinations of the following parameters: $t_{\rm cool}/t_{\rm dyn} = \{ 1, 2, 3, 5, 10 \}$,  $t_{\rm dyn}/t_{\rm rot} = \{ 0,0.25, 0.5, 1.0 \}$, $\theta = \{ 0, 5, 15, 30, 45, 60, 90\}\degree$, for a total of $110$ simulations. All the time scales are calculated at $y=1$. We stop all our simulations at $t=10 t_{\rm cool}$.

\subsection{Results}

\begin{figure*}
\centering
\includegraphics[width=0.5\textwidth]{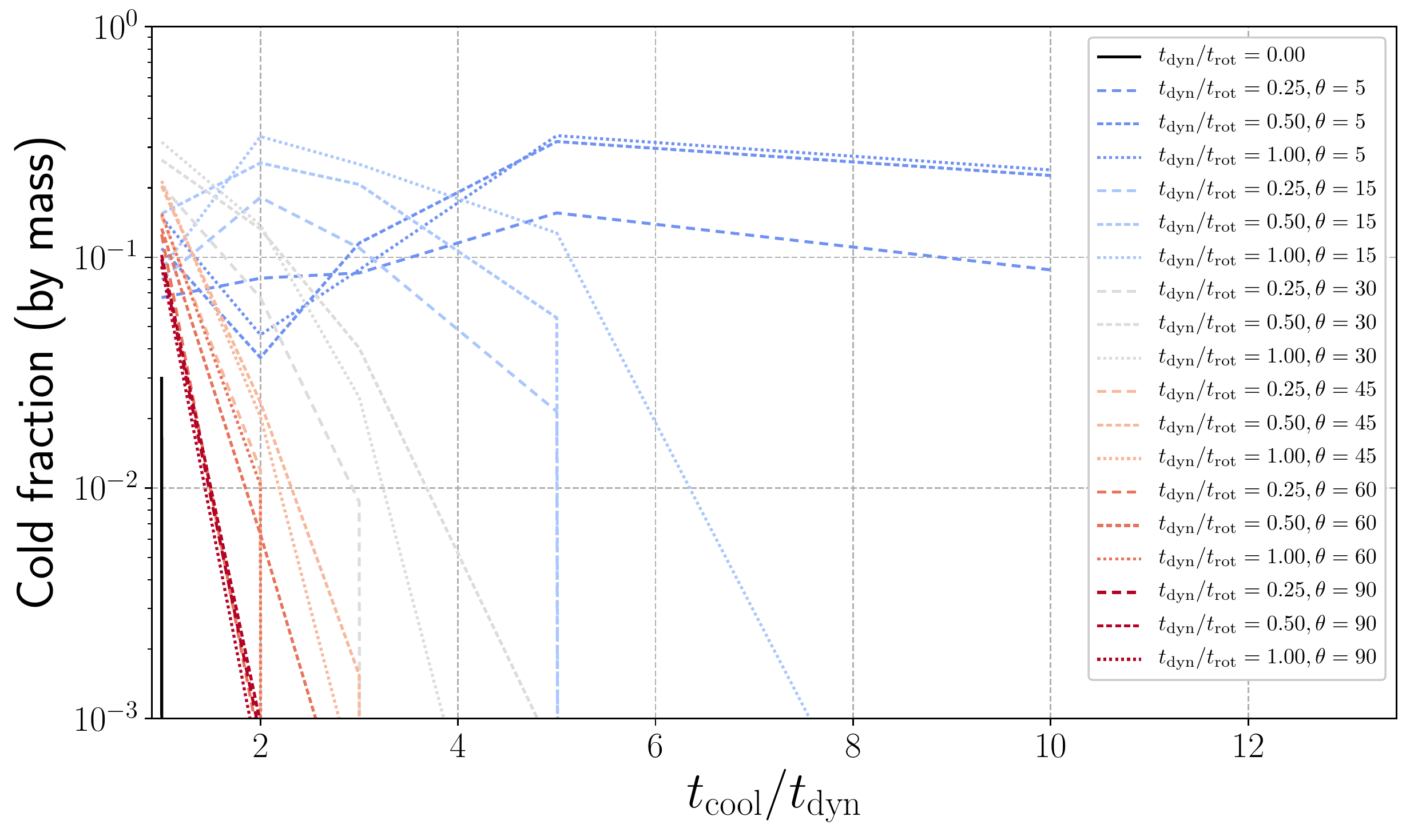}\includegraphics[width=0.5\textwidth]{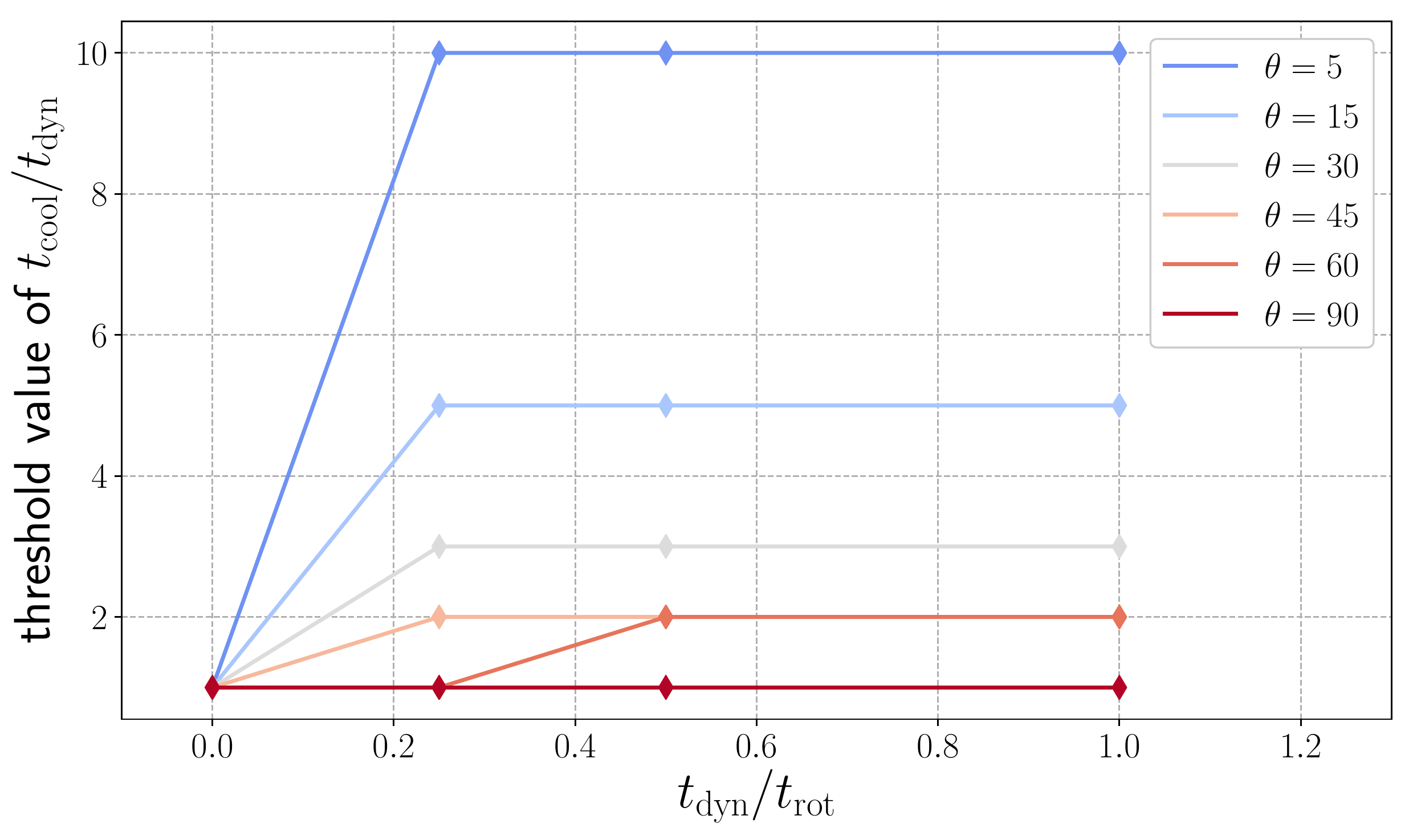}
\caption{\emph{Left:} mass fraction of cold gas (defined as gas with $T \leq T_0/3$ averaged on the region $1<y/H<3$ and at times $5<t/t_{\rm cool}<10$) as a function of $t_{\rm cool}/t_{\rm dyn}$. The threshold value of $t_{\rm cool}/t_{\rm dyn}$ above which the cold fraction drops off sharply depends on $t_{\rm dyn}/t_{\rm rot}$ and $\theta$. \emph{Right:} the threshold value of $t_{\rm cool}/t_{\rm dyn}$ (defined as cold fraction $>5\times10^{-3}$) plotted as a function of $t_{\rm dyn}/t_{\rm rot}$ for various values of $\theta$. Rotation enhances condensation by raising the threshold value below which condensation can form by a factor up to $\sim 10$.}
\label{fig:condensation}
\end{figure*}

In Figure \ref{fig:main} we show the snapshot at $t=7t_{\rm cool}$ of our simulations for various values of $t_{\rm dyn}/t_{\rm rot}$ and $t_{\rm cool}/t_{\rm dyn}$ in the case $\theta=15\degree$, i.e. for the case in which ${\pmb \Omega}$ and ${\bf g}_{\rm eff}$ are approximately perpendicular. This case resembles the typical conditions of a galactic atmosphere close to the equatorial plane.

The leftmost column corresponds to the non-rotating case ($t_{\rm dyn}/t_{\rm rot}=0$) and reproduces the results of \citet{McCourt+2012}. This shows that when the cooling time is shorter or comparable to the dynamical time ($t_{\rm cool}/t_{\rm dyn}\lesssim 1$) the thermal instability leads to the condensation of clumps, while when the cooling time is longer than the dynamical time ($t_{\rm cool}/t_{\rm dyn}>1$) condensations do not occur, in agreement with previous results \citep{McCourt+2012}.

The central and rightmost columns shows what happens when when rotation becomes dynamically important, $t_{\rm dyn}/t_{\rm rot}=0.25, 0.5$. When $t_{\rm cool}/t_{\rm dyn}\lesssim 1$ (top row), the condensation process is similar to the non-rotating case, the main difference being that mixing in the direction perpendicular to ${\pmb \Omega}$ is suppressed compared to the non-rotating case due to the conservation of angular momentum. 

The most interesting panels are the ones with $t_{\rm cool}/t_{\rm dyn} > 1$ and $t_{\rm dyn}/t_{\rm rot}=0.25, 0.5$ (the 2x2 square at the bottom-right). These show that, in contrast to the non-rotating case, condensations can still happen when $t_{\rm cool}/t_{\rm dyn}> 1$. This shows that condensations are enhanced in the presence of rotation even if $t_{\rm cool}/t_{\rm dyn}$ is kept fixed, and is the main result of this paper. If on top of this effect we also add, as mentioned in Sect. \ref{sec:timescales}, that rotation can also decrease the value of $t_{\rm cool}/t_{\rm dyn}$ compared to a non-rotating atmosphere with the same value of $g$, we see that regions of galactic atmospheres with a significant rotational support fall much more easily in the regime in which condensations can form.

Fig. \ref{fig:angle} shows the dependence of the condensations on the angle $\theta$ (see Fig. \ref{fig:sketch} for the definition of $\theta$). We see that the enhancement of the thermal instability due to rotation is maximum when $\theta=0$, i.e. when ${\pmb \Omega}$ and $\mathbf{g}_{\rm eff}$ are perpendicular (which in real systems is more likely to happen in proximity of the equatorial plane), while it is minimum (almost absent) when ${\pmb \Omega}$ and $\mathbf{g}_{\rm eff}$ are parallel.

Fig. \ref{fig:condensation} puts this on quantitative grounds. The left panel shows how the fraction of condensed (cold) gas depends on the parameters. We see that the fraction of condensed gas drops off sharply above some threshold value of $t_{\rm cool}/t_{\rm dyn}$ in all curves. The black line corresponds to the case studied by \cite{McCourt+2012} and shows that in the absence of rotation this threshold value is at $t_{\rm cool}/t_{\rm dyn} = 1$. The other lines show that the fraction of condensed gas \& the threshold value increase when either $t_{\rm dyn}/t_{\rm rot}$ is increased (more rotation) or $\theta$ is decreased (${\pmb \Omega}$ and $\mathbf{g}_{\rm eff}$ more perpendicular). 

The right panel shows the dependence of the threshold value of $t_{\rm cool}/t_{\rm dyn}$ below which condensation can form on $t_{\rm dyn}/t_{\rm rot}$ and $\theta$. It shows that the presence of rotation can rise the threshold value by a factor up to $\sim10$, depending on the value of $t_{\rm dyn}/t_{\rm rot}$ and $\theta$. The effect of rotation is important when $t_{\rm dyn}/t_{\rm rot}\gtrsim 0.2$, i.e. when rotation is dynamically important ($\Omega c_{\rm s}/g_{\rm eff}\gtrsim 1$). This regime is complementary to that studied by \citet{Gaspari+2015}, who performed global simulations focusing on relatively slow rotating atmospheres.

\subsection{Interpretation of the results} \label{sec:interpretation}

The fact that, in general, condensations are more likely to occur in the presence of rotation ($t_{\rm dyn}/t_{\rm rot}$ is large) may be naively expected from the linear stability analysis of Sect. \ref{sec:DR}, because rotation causes the presence of a new unstable mode. However, the fact that the enhancement of condensations depends on the value of $\theta$ is unexpected and more difficult to explain, because the new unstable mode (as well as any other linear mode according to the dispersion relation of Sect. \ref{sec:DR}) does \emph{not} depend on the value of $\theta$. Thus, non-linear effects must play a key role. One possible interpretation is as follows.

Even in the non-rotating case, the dispersion relation always admits a formally unstable solution growing with a rate $\omega \sim -i\omega_{\rm th}$ (see Eq. \ref{eq:mode1}). As discussed in the last paragraph Section \ref{sec:slowcooling}, in the slow cooling regime these formally unstable modes correspond to modes such that the wave vector ${\bf k}$ is nearly aligned with the effective gravity ${\bf g}_{\rm eff}$. \citet{Binney+2009} suggested that these modes are unlikely to produce condensations. Their argument is based on non-linear effects and is along the following lines. The unstable modes in which ${\bf k}$ is aligned to ${\bf g}_{\rm eff}$ correspond to modes in which blobs are moving horizontally in our setup (see Eq. \ref{eq:lin_cont}). However, as soon as blobs become slightly overdense while they cool, they start sinking and therefore moving vertically (contrary to what is predicted by the linear theory), and the subsequent mixing with the rest of the gas prevents condensations. In the parlance of linear theory (even if this effect is intrinsically non linear) one may say that the sinking effectively `changes' the ${\bf k}$ towards that of stable modes. The fact that thermal condensation does not occur when $t_{\rm cool}/t_{\rm dyn}>1$ in the non-rotating case (left column of Fig. \ref{fig:main}) is in agreement with this argument.

We argue that the above argument is no longer valid when rotation is dynamically important. If $t_{\rm dyn}/t_{\rm rot}$ is large, the motion in the direction perpendicular to ${\pmb \Omega}$ is strongly inhibited by the conservation of the angular momentum.\footnote{Indeed, the Coriolis term in Eq. \eqref{eq:euler} has the same mathematical structure of the Lorentz force in an external magnetic field, which is well-known to have the effect of aligning structures in the direction of the magnetic field.} When $\theta\sim 0$ the direction in which motion is inhibited is the vertical direction in our setup. This means that an overdense blob is prevented from sinking when it becomes slightly overdense in the above argument. By preventing an overdense blob from sinking, the conservation of angular momentum also prevents it from mixing with the rest and therefore favours condensation. By contrast, when $\theta=90\degree$ sinking is not prevented, so condensation proceeds similarly to the non-rotating case. We may re-express this argument in the parlance of the linear theory (even if this effect is intrinsically non-linear) by saying that the direction of ${\bf k}$ is `locked' onto the direction perpendicular to ${\pmb \Omega}$, so those modes will be more prominent during the subsequent non-linear evolution of the system.

\section{Relevance for astrophysical systems}
\label{sec:relevance}

The analysis of the last section suggests that the condensation of overdense clumps in rotating stratified atmospheres is more likely to occur in regions where (i) the rotation is dynamically important, and (ii) the effective gravity and the angular velocity are nearly perpendicular. In real systems, this typically happens close in the inner regions and in proximity of the equatorial plane.

As we briefly outline in the following, these results may be relevant for the interpretation of the cold structures that are often observed in the circumgalactic medium of galaxies (for example the High Velocity Clouds in the Milky Way) and in clusters (mainly cold filaments detected through their H$\alpha$ emission). Of course our idealised local analysis is only meant to be suggestive, and needs to be supported by more detailed studies of the individual cases.

\subsection{High Velocity Clouds in the Milky Way corona}

HVC are typically found in the proximity of the Galactic disc. Most of the HVC are at distances $\lesssim 15\kpc$, heights $\lesssim 10\kpc$, and within $30\degree$ from the Galactic plane as viewed from the Galactic center \citep[see e.g.][and references therein]{Putman+2012}. The observed low metallicities (typically in the range $\sim 0.1-0.3 Z_\odot$; e.g. \citealt{vanWoerdenWakker2004}) suggest that HVC are made of gas that is entering the Galaxy for the first time.

Thermal instability is a natural candidate to explain the formation of the HVC by condensation from the hot gas phase \citep[e.g.][]{MallerBullock2004, Kaufmann+2006, Peek+2008}. This possibility was criticised by \citet{Binney+2009} based on the fact that the thermal instability of cooling flows is suppressed by buoyancy and thermal conduction in non-rotating, pressure supported coronae. \citet{Sharma+2012} instead argued that HVC may form close to the virial radius of the Milky Way, according to the location of the minimum of the ratio between the cooling and the dynamical time. However, according to this picture it is unclear (i) why HVC are observed only close to the Galactic disc; (ii) how clouds formed at the virial radius manage to survive all the way down to the disc without being disrupted, for example by the Kelvin-Helmholtz instability.\footnote{Regarding the item (i), it should be noted that while all non-Magellanic HVCs are found in proximity of the Galactic plane, a significant amount of low velocity cold gas in the CGM (even close to the viral radius) may be missed by current observations \citep[e.g.][]{Zheng2015}. Regarding the item (ii), it is possible that the stripped cold gas is replenished by gas cooling from a warmer phase \citep[e.g.][]{GronkePeng2018}.}

Since the coronae of spiral galaxies are expected to rotate significantly in the inner parts \citep[e.g.][]{Oppenheimer2018}, our paper suggests that, by increasing the threshold value of $t_{\rm cool}/t_{\rm dyn}$ below which condensations can form, thermal condensations resembling HVCs might appear in those regions. If the HVC indeed form close to the disc, there is less room for the Kelvin-Helmholtz instability to disrupt them. We explore this topic in much more detail in Paper {\sc II}, in which we study specifically the case of the Galactic corona using the models presented in \citet{Sormani2018}.

\subsection{Cold filaments in the cool cores of galaxy clusters}

It has been previously suggested that cold filaments in galaxy clusters can develop only if the ratio of the cooling to the dynamical time is $t_{\rm cool}/t_{\rm dyn}\lesssim 10$ \citep[e.g.][]{Sharma+2012, Gaspari+2012}. However, recent observational studies have challenged this expectation by finding that, even if the filaments are preferentially located in the clusters with the lowest $t_{\rm cool}/t_{\rm dyn}$, the threshold is somewhat higher ($t_{\rm cool}/t_{\rm dyn}\lesssim 30$), and even the existence of clusters having $t_{\rm cool}/t_{\rm dyn}\lesssim 10$ is controversial \citep[e.g.][]{Hogan+2017, Pulido+2018}. Moreover, most of the scatter in $t_{\rm cool}/t_{\rm dyn}$ seems to be due to variations of $t_{\rm cool}$ among different clusters.

Our results show that, in the cases when the intracluster medium rotates significantly (see the discussion in the Introduction), this tension might be resolved by the fact that the threshold value of $t_{\rm cool}/t_{\rm dyn}$ below which condensations can form is increased. The right panel in Fig. \ref{fig:condensation} suggests that the required increase of a factor of $3$ can be accommodated relatively easily in a significantly rotating intracluster medium. Interestingly, \citet{Olivares+2019} pointed out that Abell 262, the cluster in their sample (out of a total of 15) in which the cold gas is condensing at the highest value of $t_{\rm cool}/t_{\rm dyn}\sim 30$ (in all other cases they found condensations at $t_{\rm cool}/t_{\rm dyn}\lesssim 20$), is also the one that shows the clearest signatures of a global rotation pattern in the CO velocity map together with Hydra-A. In general, it is reasonable to expect that $t_{\rm cool}/t_{\rm dyn}$ is shorter on the equatorial plane. Thus, it might be that for some clusters $t_{\rm cool}/t_{\rm dyn}\gtrsim 10$ when averaged, but that $t_{\rm cool}/t_{\rm dyn}$ is locally lower on the plane. 

Since filaments should form preferentially on the plane perpendicular to the rotational axis, when viewed in projection on the plane of the sky one may therefore expect to see nearly isotropically distributed filaments if the angular velocity is directed along the line of sight, and more anisotropic filaments if the angular velocity is instead perpendicular to the line of sight. This scenario may be checked observationally, assuming that the equatorial plane can be identified through the position of the cold fronts.

Finally, if the density and pressure of the hot gas are set by the requirement that the gas the value of $t_{\rm cool}/t_{\rm dyn}$ is close to the threshold value for the production of condensations \citep[e.g.][]{Sharma+2012b, Voit+2015}, this will affect the structure of a rotating atmosphere. The consequences of this requirement are yet to be explored in the context of rotating coronae.

\section{Conclusions}
\label{sec:conclusions}

We have studied the effect of rotation on the thermal stability of stratified galactic atmospheres, which include the circumgalactic medium of galaxies and the intracluster medium in galaxy clusters, using a plane parallel idealised setup. We have assumed that the system is in global thermal equilibrium, namely we have assumed that some heating source is exactly balanced by the cooling on average and on large enough spatial scales. We have neglected the effects of magnetic fields and thermal conduction since these have been studied extensively by \citet{McCourt+2012} and we wanted to focus on the effects of rotation. We have also neglected the effects of differential rotation. We have used both analytic arguments and numerical simulations.

The development of condensations via thermal instability is regulated by three fundamental time scales: (1) the dynamical time $t_{\rm dyn}$, which is the typical time scale of buoyant oscillations. This is defined as $t_{\rm dyn}=\sqrt{2}c_{\rm s}/g_{\rm eff}$, where $c_{\rm s}$ is the sound speed and ${\bf g}_{\rm eff}=-\nabla\Phi + \Omega^2{\bf R}=\nabla P/\rho$ is the local effective gravity;\footnote{As discussed in Section \ref{sec:timescales}, in the case of a plane parallel atmosphere our definition of dynamical time is equivalent to the usual free-fall time, $t_{\rm ff}=\sqrt{2H/g_{\rm eff}}$, where $H=c_{\rm s}^2/g_{\rm eff}$ is the typical height scale of the system. Our alternative definition is based only on local quantities, and it is therefore easier to generalise to the rotating, non spherically symmetric atmospheres that we consider in Paper {\sc II}.} (2) the rotational time scale $t_{\rm rot}=2\pi/\Omega$, where $\Omega$ is the angular velocity of the system; (3) the cooling time scale $t_{\rm cool}$, which is associated with the typical growth rate of the thermally unstable modes.

We have found that:
\begin{enumerate}
\item Cold structures may condense from the hot medium only if $t_{\rm cool}/t_{\rm dyn}$ is below a threshold value.
\item Condensation is enhanced in the presence of rotation. In general, this effect is important when $\Omega c_{\rm s}/g_{\rm eff}\gtrsim 1$. The fraction of condensed gas depends on the amount of rotation (parameter $t_{\rm dyn}/t_{\rm rot}$), and on the relative orientation of the rotation axis and the local effective gravity (parameter $\theta$ defined in Fig. \ref{fig:sketch}). This is quantified in the left panel of Fig. \ref{fig:condensation}. The enhancement of the thermal instability is maximum when the rotation axis and the effective gravity vector are perpendicular ($\theta=0$).

\item The threshold value of $t_{\rm cool} / t_{\rm dyn}$ below which condensations can form also depends on $t_{\rm dyn}/t_{\rm rot}$ and $\theta$. The left panel in Fig. \ref{fig:condensation} shows the dependence of the threshold on these two parameters. This is the main result of this paper. The threshold can be increased up by a factor of $\sim 10$ if rotation is significant and the rotation axis and the effective gravity vector are perpendicular ($\theta=0$).
\item These results might be relevant for the formation of HVCs (see Paper {\sc II}).
\item These results might be relevant for the formation of cold filaments in the cool cores of galaxy clusters. This has consequences on the orientations of filaments that may be observationally checked.
\end{enumerate}
The presence of rotation enhances the thermal instability in two ways. First, directly, by the effect discovered in this paper that condensation is enhanced in the presence of rotation even if $t_{\rm dyn}/t_{\rm cool}$ is kept fixed. The physical interpretation of this effect is discussed in Sect. \ref{sec:interpretation}. Second, indirectly, by reducing $|\mathbf{g}_{\rm eff}|$ and thus decreasing the ratio $t_{\rm cool}/t_{\rm dyn}$ for the same value of ${\bf g}$ (see also Section 7.1 of \citealt{Voit+2017}). The net result is that condensation may occur even in the moderately slow cooling regime if the effect of rotation is dynamically important. These conditions are more naturally met in the central parts and/or close to the equatorial plane of rotating atmospheres.

\section*{Acknowledgements}

The authors thank Filippo Fraternali, Uri Keshet, Carlo Nipoti, Gabriele Pezzulli and Steve Shore for useful comments and discussions. We are grateful to the referee, Prateek Sharma, for constructive suggestions that improved the paper. ES acknowledges support from the Israeli Science Foundation (grant 719/14) and from the German Israeli Foundation for Scientific Research and Development (grant I-1362-303.7/2016). MCS acknowledges support from the Deutsche Forschungsgemeinschaft via the Collaborative Research Centre (SFB 881) ``The Milky Way System'' (sub-projects B1, B2, and B8).

%%%%%%%%%%%%%%%%%%%%%%%%%%%%%%%%%%%%%%%%%
\def\aap{A\&A}\def\aj{AJ}\def\apj{ApJ}\def\mnras{MNRAS}\def\araa{ARA\&A}\def\aapr{Astronomy \&
 Astrophysics Review}\def\apjs{ApJS}\def\apjl{ApJ}\def\pasj{PASJ}\def\nat{Nature}\def\prd{Phys. Rev. D}
\def\ssr{Space Sci. Rev.}\def\pasp{PASP}\def\pasa{Publications of the Astronomical Society of Australia}\def\physrep{Phys. Rep.}
\bibliographystyle{mn2e}
\bibliography{bibliography}

\begin{thebibliography}{52}
\expandafter\ifx\csname natexlab\endcsname\relax\def\natexlab#1{#1}\fi

\bibitem[{{Balbus} \& {Potter}(2016)}]{BalbusPotter2016}
{Balbus} S.~A., {Potter} W.~J., 2016, Reports on Progress in Physics, 79,
  066901

\bibitem[{{Bianconi} {et~al}\mbox{.}(2013){Bianconi}, {Ettori}, \&
  {Nipoti}}]{Bianconi+2013}
{Bianconi} M., {Ettori} S., {Nipoti} C., 2013, \mnras, 434, 1565

\bibitem[{{Binney} {et~al}\mbox{.}(2009){Binney}, {Nipoti}, \&
  {Fraternali}}]{Binney+2009}
{Binney} J., {Nipoti} C., {Fraternali} F., 2009, \mnras, 397, 1804

\bibitem[{{Choudhury} \& {Sharma}(2016)}]{ChoudhurySharma2016}
{Choudhury} P.~P., {Sharma} P., 2016, \mnras, 457, 2554

\bibitem[{{Choudhury} {et~al}\mbox{.}(2019){Choudhury}, {Sharma}, \&
  {Quataert}}]{Choudhury+2019}
{Choudhury} P.~P., {Sharma} P., {Quataert} E., 2019, arXiv:1901.02903

\bibitem[{{Conselice} {et~al}\mbox{.}(2001){Conselice}, {Gallagher}, \&
  {Wyse}}]{Conselice+2001}
{Conselice} C.~J., {Gallagher}, III J.~S., {Wyse} R.~F.~G., 2001, \aj, 122,
  2281

\bibitem[{{Fabian} {et~al}\mbox{.}(2008){Fabian}, {Johnstone}, {Sanders},
  {Conselice}, {Crawford}, {Gallagher}, \& {Zweibel}}]{Fabian2008}
{Fabian} A.~C., {Johnstone} R.~M., {Sanders} J.~S., {Conselice} C.~J.,
  {Crawford} C.~S., {Gallagher}, III J.~S., {Zweibel} E., 2008, \nat, 454, 968

\bibitem[{{Field}(1965)}]{Field1965}
{Field} G.~B., 1965, \apj, 142, 531

\bibitem[{{Gaspari} {et~al}\mbox{.}(2015){Gaspari}, {Brighenti}, \&
  {Temi}}]{Gaspari+2015}
{Gaspari} M., {Brighenti} F., {Temi} P., 2015, \aap, 579, A62

\bibitem[{{Gaspari} {et~al}\mbox{.}(2012){Gaspari}, {Ruszkowski}, \&
  {Sharma}}]{Gaspari+2012}
{Gaspari} M., {Ruszkowski} M., {Sharma} P., 2012, \apj, 746, 94

\bibitem[{{Gronke} \& {Oh}(2018)}]{GronkePeng2018}
{Gronke} M., {Oh} S.~P., 2018, \mnras, 480, L111

\bibitem[{{Hodges-Kluck} {et~al}\mbox{.}(2016){Hodges-Kluck}, {Miller}, \&
  {Bregman}}]{HodgesKluck+2016}
{Hodges-Kluck} E.~J., {Miller} M.~J., {Bregman} J.~N., 2016, \apj, 822, 21

\bibitem[{{Hogan} {et~al}\mbox{.}(2017){Hogan}, {McNamara}, {Pulido}, {Nulsen},
  {Vantyghem}, {Russell}, {Edge}, {Babyk}, {Main}, \& {McDonald}}]{Hogan+2017}
{Hogan} M.~T. {et~al.}, 2017, \apj, 851, 66

\bibitem[{{Joung} {et~al}\mbox{.}(2012){Joung}, {Bryan}, \&
  {Putman}}]{Joung+2012}
{Joung} M.~R., {Bryan} G.~L., {Putman} M.~E., 2012, \apj, 745, 148

\bibitem[{{Kaufmann} {et~al}\mbox{.}(2006){Kaufmann}, {Mayer}, {Wadsley},
  {Stadel}, \& {Moore}}]{Kaufmann+2006}
{Kaufmann} T., {Mayer} L., {Wadsley} J., {Stadel} J., {Moore} B., 2006, \mnras,
  370, 1612

\bibitem[{{Keshet} {et~al}\mbox{.}(2010){Keshet}, {Markevitch}, {Birnboim}, \&
  {Loeb}}]{Keshet+2010}
{Keshet} U., {Markevitch} M., {Birnboim} Y., {Loeb} A., 2010, \apjl, 719, L74

\bibitem[{{Malagoli} {et~al}\mbox{.}(1987){Malagoli}, {Rosner}, \&
  {Bodo}}]{Malagoli+1987}
{Malagoli} A., {Rosner} R., {Bodo} G., 1987, \apj, 319, 632

\bibitem[{{Maller} \& {Bullock}(2004)}]{MallerBullock2004}
{Maller} A.~H., {Bullock} J.~S., 2004, \mnras, 355, 694

\bibitem[{{Manolopoulou} \& {Plionis}(2017)}]{ManolopoulouPlionis2017}
{Manolopoulou} M., {Plionis} M., 2017, \mnras, 465, 2616

\bibitem[{{Markevitch} \& {Vikhlinin}(2007)}]{MarkevitchVikhlinin2007}
{Markevitch} M., {Vikhlinin} A., 2007, \physrep, 443, 1

\bibitem[{{McCourt} {et~al}\mbox{.}(2012){McCourt}, {Sharma}, {Quataert}, \&
  {Parrish}}]{McCourt+2012}
{McCourt} M., {Sharma} P., {Quataert} E., {Parrish} I.~J., 2012, \mnras, 419,
  3319

\bibitem[{{McDonald} {et~al}\mbox{.}(2011){McDonald}, {Veilleux}, \&
  {Mushotzky}}]{McDonald2011}
{McDonald} M., {Veilleux} S., {Mushotzky} R., 2011, \apj, 731, 33

\bibitem[{{McDonald} {et~al}\mbox{.}(2010){McDonald}, {Veilleux}, {Rupke}, \&
  {Mushotzky}}]{McDonald2010}
{McDonald} M., {Veilleux} S., {Rupke} D.~S.~N., {Mushotzky} R., 2010, \apj,
  721, 1262

\bibitem[{{McNamara} {et~al}\mbox{.}(2014){McNamara}, {Russell}, {Nulsen},
  {Edge}, {Murray}, {Main}, {Vantyghem}, {Combes}, {Fabian}, {Salome},
  {Kirkpatrick}, {Baum}, {Bregman}, {Donahue}, {Egami}, {Hamer}, {O'Dea},
  {Oonk}, {Tremblay}, \& {Voit}}]{McNamara+2014}
{McNamara} B.~R. {et~al.}, 2014, \apj, 785, 44

\bibitem[{{Mignone} {et~al}\mbox{.}(2007){Mignone}, {Bodo}, {Massaglia},
  {Matsakos}, {Tesileanu}, {Zanni}, \& {Ferrari}}]{Pluto2007}
{Mignone} A., {Bodo} G., {Massaglia} S., {Matsakos} T., {Tesileanu} O., {Zanni}
  C., {Ferrari} A., 2007, \apjs, 170, 228

\bibitem[{{Nipoti}(2010)}]{Nipoti2010}
{Nipoti} C., 2010, \mnras, 406, 247

\bibitem[{{Nipoti} \& {Posti}(2014)}]{NipotiPosti2014}
{Nipoti} C., {Posti} L., 2014, \apj, 792, 21

\bibitem[{{Nipoti} {et~al}\mbox{.}(2015){Nipoti}, {Posti}, {Ettori}, \&
  {Bianconi}}]{Nipoti+2015}
{Nipoti} C., {Posti} L., {Ettori} S., {Bianconi} M., 2015, Journal of Plasma
  Physics, 81, 495810508

\bibitem[{{Olivares} {et~al}\mbox{.}(2019){Olivares}, {Salom{\'e}}, {Combes},
  {Hamer}, {Guillard}, {Lehnert}, {Polles}, {Beckmann}, {Dubois}, {Donahue},
  {Edge}, {Fabian}, {McNamara}, {Rose}, {Russell}, {Tremblay}, {Vantyghem},
  {Canning}, {Ferland}, {Godard}, {Hogan}, {Peirani}, \& {Pineau des
  Forets}}]{Olivares+2019}
{Olivares} V. {et~al.}, 2019, arXiv:1902.09164

\bibitem[{{Oppenheimer}(2018)}]{Oppenheimer2018}
{Oppenheimer} B.~D., 2018, \mnras, 480, 2963

\bibitem[{{Peek} {et~al}\mbox{.}(2008){Peek}, {Putman}, \&
  {Sommer-Larsen}}]{Peek+2008}
{Peek} J.~E.~G., {Putman} M.~E., {Sommer-Larsen} J., 2008, \apj, 674, 227

\bibitem[{{Pulido} {et~al}\mbox{.}(2018){Pulido}, {McNamara}, {Edge}, {Hogan},
  {Vantyghem}, {Russell}, {Nulsen}, {Babyk}, \& {Salom{\'e}}}]{Pulido+2018}
{Pulido} F.~A. {et~al.}, 2018, \apj, 853, 177

\bibitem[{{Putman} {et~al}\mbox{.}(2012){Putman}, {Peek}, \&
  {Joung}}]{Putman+2012}
{Putman} M.~E., {Peek} J.~E.~G., {Joung} M.~R., 2012, \araa, 50, 491

\bibitem[{{Russell} {et~al}\mbox{.}(2019){Russell}, {McNamara}, {Fabian},
  {Nulsen}, {Combes}, {Edge}, {Madar}, {Olivares}, {Salome}, \&
  {Vantyghem}}]{Russell+2019}
{Russell} H.~R. {et~al.}, 2019, arXiv:1902.09227

\bibitem[{{Salom{\'e}} \& {Combes}(2004)}]{SalomeCombes2004}
{Salom{\'e}} P., {Combes} F., 2004, \aap, 415, L1

\bibitem[{{Salom{\'e}} {et~al}\mbox{.}(2006){Salom{\'e}}, {Combes}, {Edge},
  {Crawford}, {Erlund}, {Fabian}, {Hatch}, {Johnstone}, {Sanders}, \&
  {Wilman}}]{Salome+2006}
{Salom{\'e}} P. {et~al.}, 2006, \aap, 454, 437

\bibitem[{{Sanderson} {et~al}\mbox{.}(2009){Sanderson}, {O'Sullivan}, \&
  {Ponman}}]{Sanderson2009}
{Sanderson} A.~J.~R., {O'Sullivan} E., {Ponman} T.~J., 2009, \mnras, 395, 764

\bibitem[{{Sanderson} {et~al}\mbox{.}(2006){Sanderson}, {Ponman}, \&
  {O'Sullivan}}]{Sanderson2006}
{Sanderson} A.~J.~R., {Ponman} T.~J., {O'Sullivan} E., 2006, \mnras, 372, 1496

\bibitem[{{Sharma} {et~al}\mbox{.}(2012{\natexlab{a}}){Sharma}, {McCourt},
  {Parrish}, \& {Quataert}}]{Sharma+2012b}
{Sharma} P., {McCourt} M., {Parrish} I.~J., {Quataert} E., 2012{\natexlab{a}},
  \mnras, 427, 1219

\bibitem[{{Sharma} {et~al}\mbox{.}(2012{\natexlab{b}}){Sharma}, {McCourt},
  {Quataert}, \& {Parrish}}]{Sharma+2012}
{Sharma} P., {McCourt} M., {Quataert} E., {Parrish} I.~J., 2012{\natexlab{b}},
  \mnras, 420, 3174

\bibitem[{{Sormani} \& {Sobacchi}(2019)}]{SormaniSobacchi2019}
{Sormani} M.~C., {Sobacchi} E., 2019, in preparation (Paper II)

\bibitem[{{Sormani} {et~al}\mbox{.}(2018){Sormani}, {Sobacchi}, {Pezzulli},
  {Binney}, \& {Klessen}}]{Sormani2018}
{Sormani} M.~C., {Sobacchi} E., {Pezzulli} G., {Binney} J., {Klessen} R.~S.,
  2018, \mnras, 481, 3370

\bibitem[{{Tumlinson} {et~al}\mbox{.}(2017){Tumlinson}, {Peeples}, \&
  {Werk}}]{Tumlinson+2017}
{Tumlinson} J., {Peeples} M.~S., {Werk} J.~K., 2017, \araa, 55, 389

\bibitem[{{van Woerden} \& {Wakker}(2004)}]{vanWoerdenWakker2004}
{van Woerden} H., {Wakker} B.~P., 2004, in Astrophysics and Space Science
  Library, Vol. 312, High Velocity Clouds, {van Woerden} H., {Wakker} B.~P.,
  {Schwarz} U.~J., {de Boer} K.~S., eds., p. 195

\bibitem[{{Vikhlinin} {et~al}\mbox{.}(2006){Vikhlinin}, {Kravtsov}, {Forman},
  {Jones}, {Markevitch}, {Murray}, \& {Van Speybroeck}}]{Vikhlinin2006}
{Vikhlinin} A., {Kravtsov} A., {Forman} W., {Jones} C., {Markevitch} M.,
  {Murray} S.~S., {Van Speybroeck} L., 2006, \apj, 640, 691

\bibitem[{{Voit} {et~al}\mbox{.}(2015){Voit}, {Donahue}, {Bryan}, \&
  {McDonald}}]{Voit+2015}
{Voit} G.~M., {Donahue} M., {Bryan} G.~L., {McDonald} M., 2015, \nat, 519, 203

\bibitem[{{Voit} {et~al}\mbox{.}(2017){Voit}, {Meece}, {Li}, {O'Shea}, {Bryan},
  \& {Donahue}}]{Voit+2017}
{Voit} G.~M., {Meece} G., {Li} Y., {O'Shea} B.~W., {Bryan} G.~L., {Donahue} M.,
  2017, \apj, 845, 80

\bibitem[{{Werk} {et~al}\mbox{.}(2013){Werk}, {Prochaska}, {Thom}, {Tumlinson},
  {Tripp}, {O'Meara}, \& {Peeples}}]{Werk2013}
{Werk} J.~K., {Prochaska} J.~X., {Thom} C., {Tumlinson} J., {Tripp} T.~M.,
  {O'Meara} J.~M., {Peeples} M.~S., 2013, \apjs, 204, 17

\bibitem[{{Werk} {et~al}\mbox{.}(2014){Werk}, {Prochaska}, {Tumlinson},
  {Peeples}, {Tripp}, {Fox}, {Lehner}, {Thom}, {O'Meara}, {Ford}, {Bordoloi},
  {Katz}, {Tejos}, {Oppenheimer}, {Dav{\'e}}, \& {Weinberg}}]{Werk2014}
{Werk} J.~K. {et~al.}, 2014, \apj, 792, 8

\bibitem[{{Werner} {et~al}\mbox{.}(2019){Werner}, {McNamara}, {Churazov}, \&
  {Scannapieco}}]{Werner+2019}
{Werner} N., {McNamara} B.~R., {Churazov} E., {Scannapieco} E., 2019, \ssr,
  215, 5

\bibitem[{{Westmeier}(2018)}]{Westmeier2018}
{Westmeier} T., 2018, \mnras, 474, 289

\bibitem[{{Zheng} {et~al}\mbox{.}(2015){Zheng}, {Putman}, {Peek}, \&
  {Joung}}]{Zheng2015}
{Zheng} Y., {Putman} M.~E., {Peek} J.~E.~G., {Joung} M.~R., 2015, \apj, 807,
  103

\end{thebibliography}

\end{document}